Icarus
\documentclass[twocolumn]{emulateapj}

\newcommand{\Msun}{\ensuremath{M_{\odot}}}
\newcommand{\Rsun}{\ensuremath{R_{\odot}}}
\newcommand{\Mdot}{\ensuremath{\dot{M}}}

\shorttitle{Detections of trans-Neptunian ice}
\shortauthors{McClure et al.}

\begin{document}

\title{Detections of trans-Neptunian ice in protoplanetary disks}


\author{M. K. McClure\altaffilmark{1,7}, C. Espaillat\altaffilmark{2}, N. Calvet\altaffilmark{1}, E. Bergin\altaffilmark{1}, P. D'Alessio\altaffilmark{3}, D. M. Watson\altaffilmark{4}, P. Manoj\altaffilmark{5}, B. Sargent\altaffilmark{6}, L. I. Cleeves\altaffilmark{1}}

\altaffiltext{1}{Department of Astronomy, The University of Michigan, 500 Church St., 830 Dennison Bldg., Ann Arbor, MI 48109; melisma@umich.edu, ncalvet@umich.edu, ebergin@umich.edu,cleeves@umich.edu}
\altaffiltext{2}{Department of Astronomy, Boston University, 725 Commonwealth Avenue, Boston, MA 02215; cce@bu.edu}
\altaffiltext{3}{Centro de Radioastronom\'{i}a y Astrof\'{i}sica, Universidad NacionalAUt\'{o}noma de M\'{e}xico, 58089 Morelia, Michoac\'{a}n, M\'{e}xico; p.dalessio@crya.unam.mx}
\altaffiltext{4}{Department of Physics and Astronomy, University of Rochester, Rochester, NY 14627, USA; dmw@pas.rochester.edu}
\altaffiltext{5}{Tata Institute of Fundamental Research, Homi Bhabha Road, Colaba, Mumbai 400005, India; manoj.puravankara@tifr.res.in}
\altaffiltext{6}{Center for Imaging Science and Laboratory for Multiwavelength Astrophysics, Rochester Institute of Technology, 54 Lomb Memorial Drive, Rochester, NY 14623, USA; baspci@rit.edu}
\altaffiltext{7}{NSF Graduate Research Fellow}

\begin{abstract}
We present {\it Herschel} Space Observatory\footnote{Herschel is an ESA space observatory with science instruments provided by European-led Principal Investigator consortia and with important participation from NASA.}  PACS spectra of T Tauri stars, in which we detect amorphous and crystalline water ice features.  Using irradiated accretion disk models, we determine the disk structure and ice abundance in each of the systems.  Combining a model-independent comparison of the ice feature strength and disk size with a detailed analysis of the model ice location, we estimate that the ice emitting region is at disk radii $>$30AU, consistent with a proto-Kuiper belt.  Vertically, the ice emits most below the photodesorption zone, consistent with Herschel observations of cold water vapor. The presence of crystallized water ice at a disk location a) colder than its crystallization temperature and b) where it should have been re-amorphized in $\sim$1 Myr suggests that localized generation is occurring; the most likely cause appears to be micrometeorite impact or planetesimal collisions.  Based on simple tests with UV models and different ice distributions, we suggest that the SED shape from 20 to 50$\mu$m may probe the location of the water ice snow line in the disk upper layers.  This project represents one of the first extra-solar probes of the spatial structure of the cometary ice reservoir thought to deliver water to terrestrial planets.

\end{abstract}


\section{Introduction}

As the most abundant non-refractory solid-state component in protoplanetary disks, water ice plays a key role in the dynamics and chemistry of planet formation.  Its sticking properties may enhance the efficiency of dust grain growth and settling \citep{okuzumi+12a}, which is important for planetesimal growth.  Water is one of the main molecular oxygen carriers, so its ice desorption fronts contribute strongly to the radial variation of disks' C/O in the gas phase and on grains \citep{oberg+11e}.  The resulting C/O ratio in the atmospheres of giant planets affects these planets' spectral signatures, which are typically hydrocarbon or water vapor absorption features \citep{madhusudhan12b}. Delivery of ice from disk reservoirs by comet or asteroid impacts is likely the origin of Earth's water \citep{hartogh+11,alexander+12}. Constraining the location of water ice in disks would enhance our understanding of all these topics.

The distribution of water ice throughout the disk is dominated by thermal desorption (sublimation), with secondary effects from photodesorption and settling.  In the disk's upper layers where most of the stellar radiation is deposited, icy grains are directly exposed to the far ultra-violet (FUV) radiation field that, for low mass stars, originates mostly in the stellar accretion shock \citep{cg98}. Water molecules should be photodesorbed from the grain surface into the gas phase \citep{pollack+94,oberg+09c}, producing water vapor even at locations in the disk with temperatures below the water ice sublimation temperature \citep{dominik+05}.  However, recent observations of water vapor emission from the DM Tau and TW Hya disks with the {\it Herschel} Space Observatory show that the water lines are weaker than predicted by chemical models including photodesorption of ice \citep{bergin+10, hogerheijde+11}.  The lower water vapor abundance in TW Hya may be due to growth and subsequent settling of the icy grains from the upper layers, which would reduce the amount of ice available to photodesorb.


The structure of water ice traces its past thermal history, because it crystallizes in an irreversible reaction at temperatures of 110 to 130K \citep{smith+94}. Protostellar infall chemical models suggest that water ice is formed amorphous in a star's natal cloud, and the bulk of the ice beyond $\sim$30AU may reach the disk without exceeding this crystallization temperature \citep{visser+09a,visser+11}. This prediction is consistent with most previous observations of 3$\mu$m water ice in the upper layers of edge-on disks \citep{pontoppidan+05, terada+07}; however, Kuiper belt objects (KBOs) and satellites of Solar system gas giants show predominately crystalline ice signatures \citep{jewitt_luu04, grundy+06}. Several mechanisms exist that can  crystalize and/or amorphize water ice in disks: desorption and recondensation onto grains \citep{kouchi+94,ciesla14a}, irradiation by high energy particles \citep{moore+92}, and collisions between large particles or planetesimals \citep{porter+10}.  A handful of crystalline ice detections at 3$\mu$m \citep{schegerer_wolf10, terada_tokunaga12} in T Tauri disks as well as 43 and 63$\mu$m with {\it ISO} in Herbig AeBe disks \citep{malfait+98b, malfait+99, creech-eakman+02} represent the missing link between primitive ices from molecular clouds and thermally processed ices in mature solar systems. In our previous paper \citep{mcclure+12}, we used {\it Herschel} to observe the 63$\mu$m crystalline ice feature for the first time in a T Tauri disk and suggested that the crystalline water ice could be the result of planetesimal collisions in the outer disk.  

Here we present {\it Herschel} PACS spectra of disks around T Tauri stars in which we detect signatures of crystalline ice and amorphous ice (Sections \ref{obsred} and \ref{obsresults}).  Through comparison of their spectral energy distributions (SEDs) with synthetic spectra from disk structure models (Section \ref{sedmods}), we constrain the radial and vertical ice emitting region (Section \ref{emit}) and conclude that we have detected a proto-Kuiper belt or other trans-Neptunian water ice reservoir (Section \ref{rdice}) below the photodesorption layer (Section \ref{uvresults}), where the crystalline ice is regenerated through impacts between dust grains and/or planetestimals (Section \ref{discussion}).

\section{Sample selection, observations, and data reduction}
\label{obsred}
The four disks in our sample, AA Tau, DO Tau, Haro 6-13, and VW Cha, were drawn from two larger {\it Herschel} programs, OT1\textunderscore mmcclure\textunderscore 1 (PI: McClure) and OT1\textunderscore cespaill\textunderscore 2 (PI: Espaillat), and selected on the basis of their water ice detections, with one non-detection included as a control case (VW Cha).  We note that our combined Herschel programs observed 50 circumstellar disks; of these, 86\% did not show prominent signs of ice.  The disks presented in this paper and \citet{mcclure+12} represent the best detections. Three of the disks are in Taurus, and one is in Cha I. The {\it Herschel} OBSIDs, {\it Spitzer} AORs, and dates of observation are given in Table \ref{sampletab}. For absolute photometric verification of these spectra and for the analysis in Section \ref{sedmods}, we gathered photometry between 0.3 and 3000$\mu$m from the literature to construct SEDs.  The references for these photometry are listed in Table \ref{reftab}. The complete SEDs, including the spectra described below, are shown in Figure \ref{rawseds}.  
 
\subsection{{\it Herschel} PACS}
We observed the sample using the PACS \citep{poglitsch+10} instrument on {\it Herschel} \citep{pilbratt+10} in range spectroscopy modes B2A (51-73$\mu$m) and R1S (102-145$\mu$m) at Nyquist-sampling ($\lambda/\Delta\lambda\sim$1500).  VW Cha was also observed with B2B (71-105$\mu$m) and R1L (139-210$\mu$m).  The data were reduced using the HIPE version 9.0 \citep{ott2010} standard data reduction pipeline, including the RSRF.  The instrument is a coarse IFU with 5$\times$5 spaxels; spectra were extracted from each one individually.  After confirming that the sources were point-source like and well-centered on the central spaxel, we applied the PSF correction to the spectrum from the central spaxel.  The uncertainty in PACS absolute flux calibration can be up to 30\%.  As the data had been observed with Nyquist sampling, which provided much greater spectral resolution than necessary at the expense of signal/noise, we down-sampled the data using two approaches.  Before calculating equivalent widths for the analysis, we median-filtered the spectra with sliding bins of 50 resolution elements, which removed the bulk of the [O I] emission line at 63$\mu$m. Therefore the [O I] line in DO Tau and AA Tau \citep{howard+13} does not contribute to the broad shape of their B2A spectra. To estimate the uncertainty in the spectral shape we did the following.  First we median filtered the spectrum with a sliding window of 4 resolution elements, to remove any line emission.  We then rebinned the flux into 1$\mu$m segments, with the rebinned flux being the average flux in that segment and the uncertainty being the standard deviation of the point-to-point variation in the 1$\mu$m bin. This method shows that although the spectra are noisy, the shape of the overall continuum is robust.

 \begin{deluxetable}{ccccc}
\tabletypesize{\small}   
\tablewidth{0pt}
\tablecaption{Observations}
\tablehead{\colhead{Star} & \colhead{{\it Herschel}} & \colhead{date} & \colhead{{\it Spitzer}} & \colhead{date} \\
	& \colhead{OBSID} &	 & \colhead{AOR} &	}
\startdata
AA Tau	&	1342240152	&	2012-02-17	&  3537152	& 2004-02-28  	\\
Haro 6-13	&	1342239763	&	2012-02-29	&  3541504	& 2004-02-28  	\\
DO Tau	&	1342240156	&	2012-02-17	&   3533056	& 2004-02-29	\\
VW Cha	&	1342233472	&	2011-12-02	&  12696832	& 2005-07-12 	\\
   		&	1342233471	&	2011-12-02	&  27066112	& 2008-06-02	
\enddata
\label{sampletab}
\end{deluxetable}

\begin{figure*}
\centering{\includegraphics[angle=0, scale=0.75]{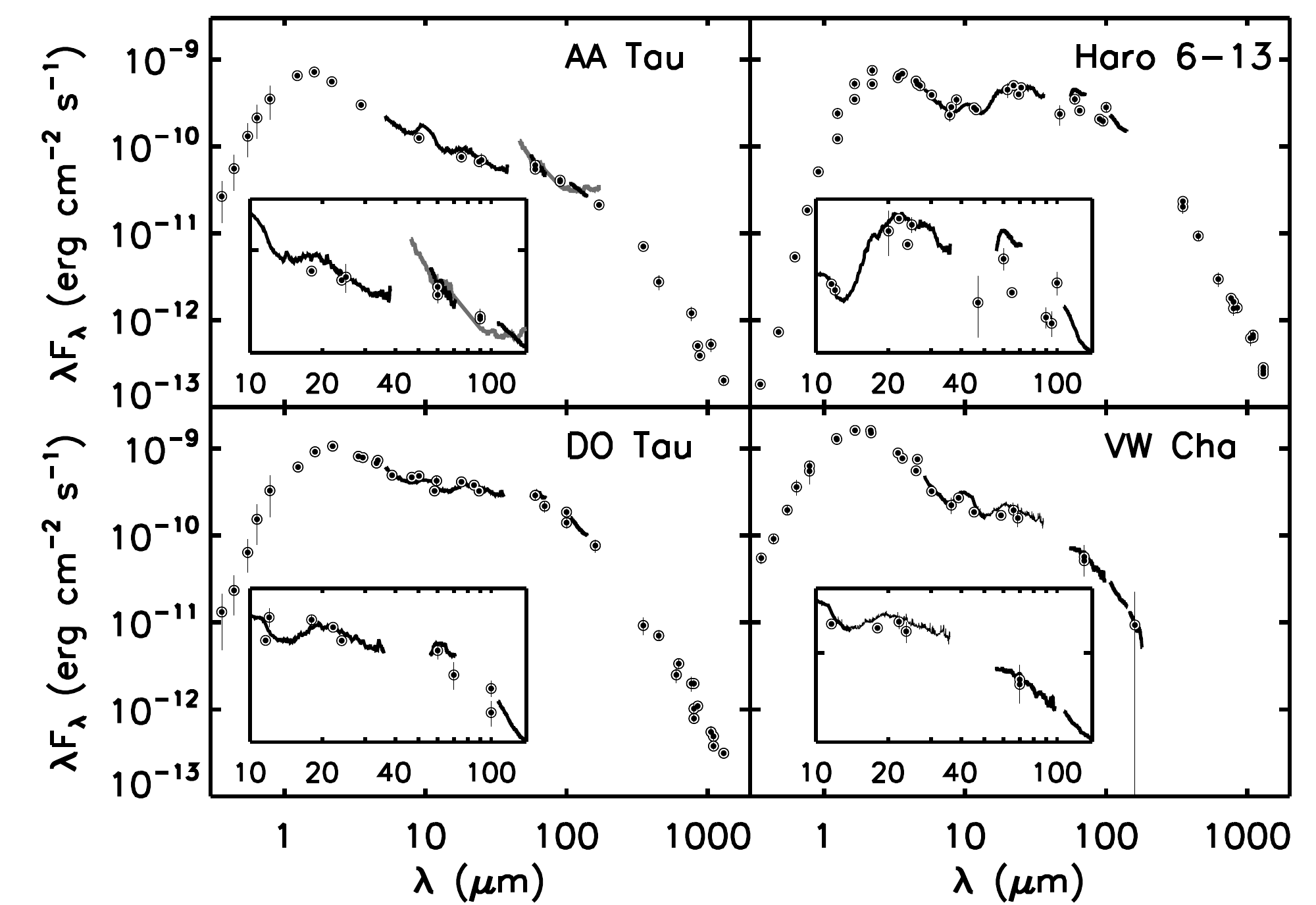}}
\caption{SEDs of targets in sample, not corrected for extinction. References for photometry are given in Table \ref{reftab}. Note that for AA Tau we also plot its {\it ISO} LWS spectrum (grey line), described in \S\ref{iso}.  \label{rawseds}}
\end{figure*}

To confirm the absolute photometric accuracy of the PACS spectrum, we compared the spectra with photometry from IRAS, AKARI, MIPS, and PACS when available.  The photometry and spectra are shown with additional data in Figure \ref{rawseds}. For AA Tau there is excellent agreement between IRAS, AKARI, ISO photometry, and the PACS spectrum.  The match between photometry, in this case MIPS and PACS, and our PACS spectroscopy is also within 1$\sigma$ for VW Cha.  The other two disks, DO Tau and Haro 6-13, show more variation between their 70$\mu$m photometry and spectroscopy, although they are nominally within 3$\sigma$ of each other after taking into account the 30\% absolute flux uncertainty in PACS. Unfortunately, the detectors used for AKARI and MIPS 70$\mu$m fluxes are optimized for sensitivity and become non-linear for source fluxes greater than $\sim$1 Jy.  Since both of these disks are $\sim$4-6 Jy at 70$\mu$m, the AKARI and MIPS fluxes only provide lower limits to the actual flux. 

\begin{deluxetable*}{cl}-eps-converted-to.pdf
\tabletypesize{\scriptsize}   
\tablewidth{0pt}
\tablecaption{Ancillary photometry}
\tablehead{\colhead{Wavelength} & \colhead{Instrument \& references}}
\startdata
UBVRI/ugriz	&	 \citet{kh95,herbst+94,sdss+12}	\\
JHK		&	2MASS \citep{cutri+03}, DENIS, \citet{myers+87}	\\
mid-IR	&	WISE \citep{cutri+12}, AKARI IRC \citep{ishihara+10}, {\it Spitzer} IRAC \citep{luhman+08} \\
far-IR	&	PACS \citep{winston+12,howard+13}, AKARI FIS, {\it Spitzer} MIPS, ISO, IRAS \\
submm  	&	\citet{aw05}, \citet{aw08}, \citet{bs91}	\\
mm		&	\citet{bs91}, \citet{beckwith+90}, \citet{kitamura+02}, \citet{dutrey+96}
\enddata
\label{reftab}
\end{deluxetable*}

\subsection{{\it Spitzer} IRS}
The {\it Spitzer} IRS \citep{houck+04} spectra were taken in some combination of low- and high-resolution modes: SL (5$-$14$\mu$m) and LL (14$-$38$\mu$m) with $\lambda/\Delta\lambda$=60$-$120 and SH (10$-$19$\mu$m) and LH (19$-$35$\mu$m) with $\lambda/\Delta\lambda$=600.  Observation dates and identifying information are given in Table \ref{sampletab}.  We reduced these data with SMART \citep{higdon+04} in the same way as in \citet{mcclure+10}, with the exception that the SH/LH data were sky subtracted from off-source frames included in their AORs.  We estimate the spectrophotometric uncertainty to be $\sim$5\%.  

\subsection{{\it ISO} LWS}
\label{iso}
One disk in our sample, AA Tau, was observed with the LWS instrument on {\it ISO} \citep{kessler+96}.  The resulting spectrum was published and analyzed by \citet{creech-eakman+02}.  We downloaded the spectrum from the ISO Data Archive as well as the reduced PHT 22 photometry at 90 and 170$\mu$m.  The LWS spectrum had several relative order mismatches and an overall offset in absolute flux from the photometry.  We chose LW2 (central $\lambda \sim$122$\mu$m) as the anchor order and scaled the other orders to that one, using the overlap between orders. We then scaled the entire spectrum down to the ISO photometry and trimmed the spectrum shortward of 46$\mu$m, between 70 and 90$\mu$m, and longward of 170$\mu$m.  The final spectrum was median filtered within sliding windows of 100 spectral elements.

\begin{figure}
\centering{\includegraphics[angle=0, scale=0.7]{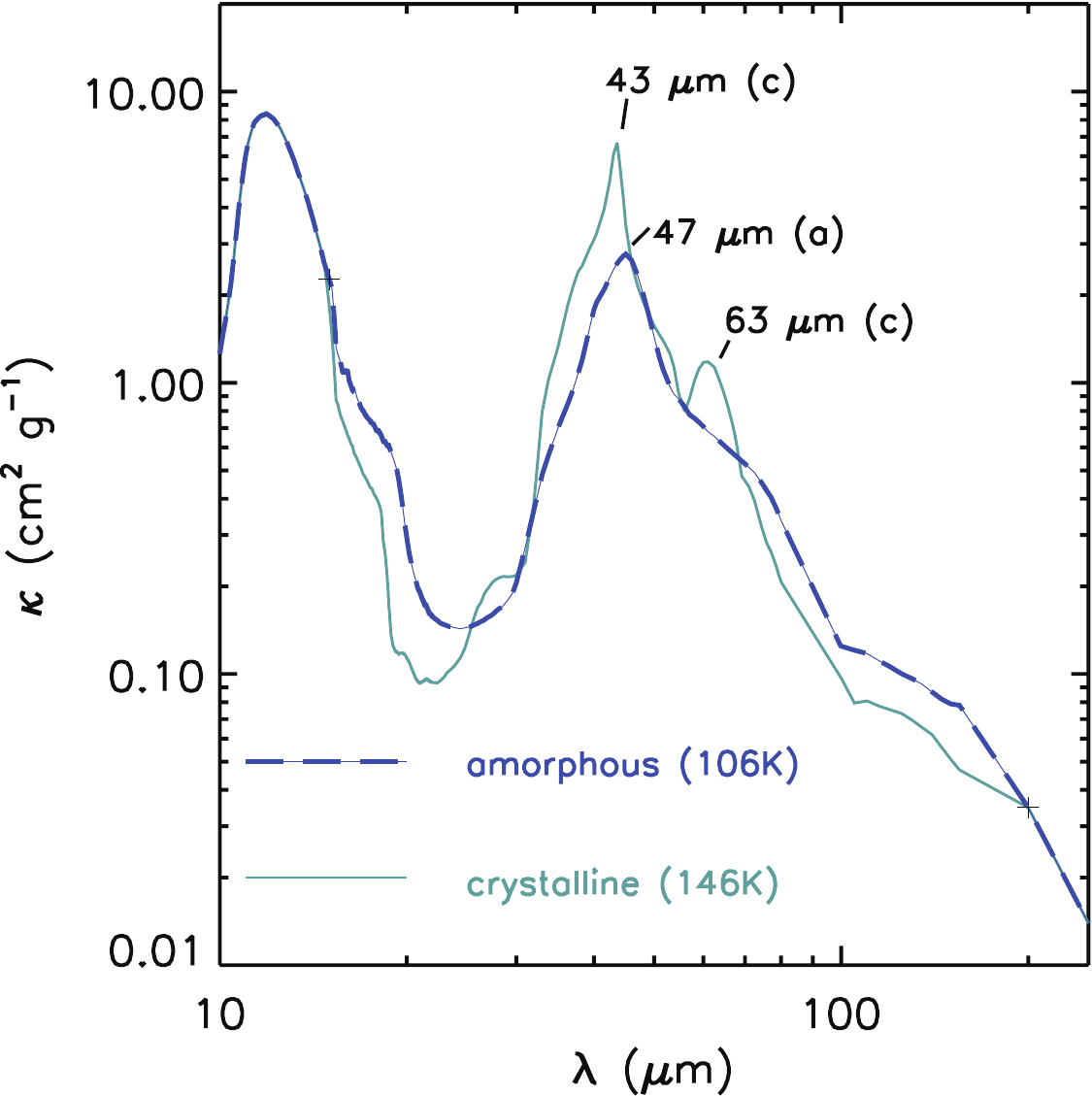}}
\caption{Detail of the far-infrared region of the hybrid water ice opacities used in the models.  Plus symbols denote the trim boundaries for the inclusion of the \citet{curtis+05} amorphous (dashed line) and crystalline (solid line) ice opacities.  Characteristic wavelengths of the far-infrared ice emission maxima are noted (with c for the crystalline features and a for the amorphous features). These opacities are for a power law grain size distribution with a maximum grain size of 0.25$\mu$m, power of -3.5, and a mass fraction of ice, relative to gas, of 0.002.  \label{wateropas}}
\end{figure}

\begin{figure*}
\centering{\includegraphics[angle=0, scale=0.74]{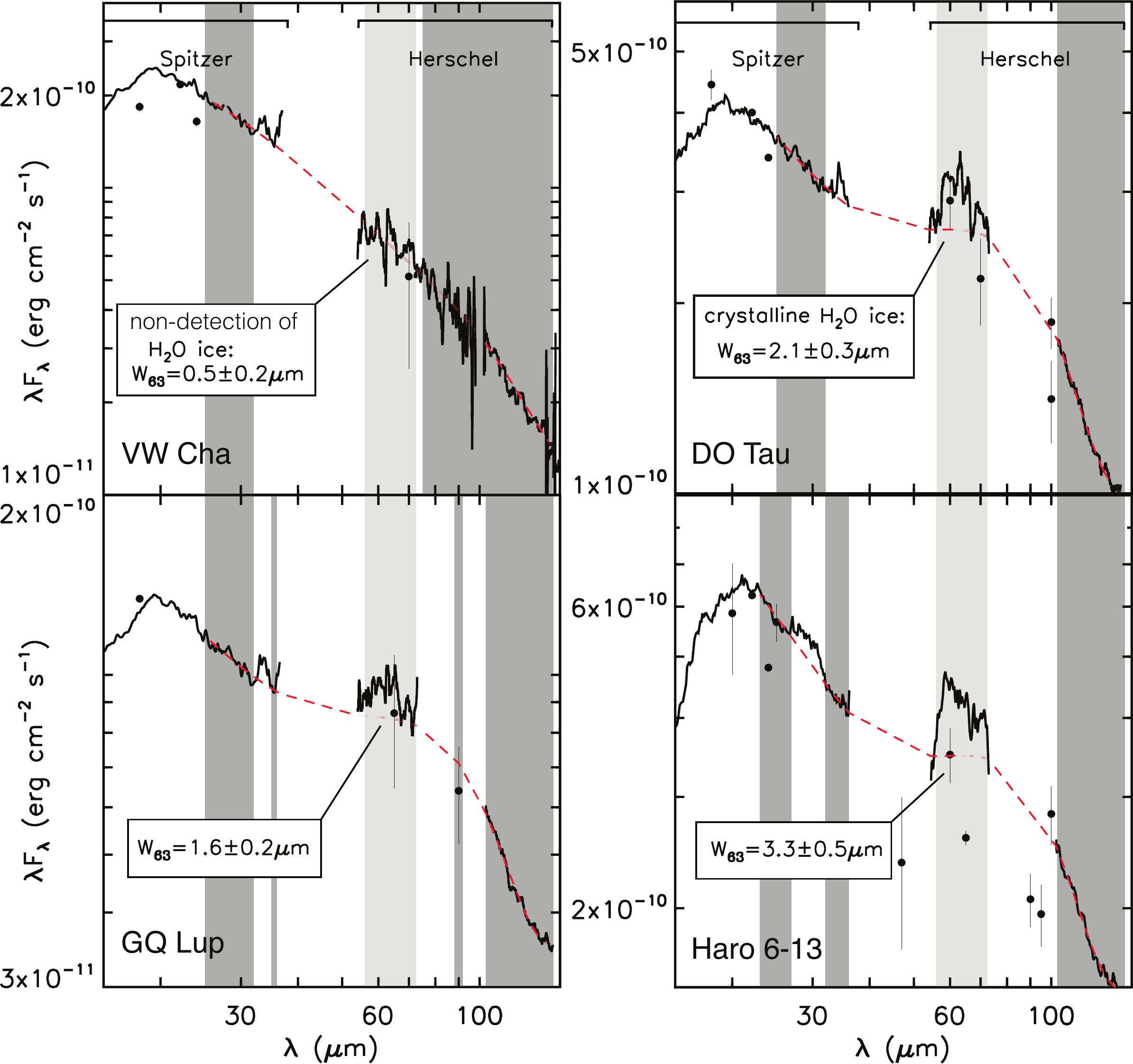}}
\caption{Determination of $W_{63}$ for the three disks in this sample, plus GQ Lup from \citet{mcclure+12}.  Continuum regions (dark gray fill), continuum fit (red dashed line), and limits of integration (light grey fill) are indicated. Error bars in $W_{63}$ are 1$\sigma$ uncertainties. \label{ew63}} \end{figure*}

\section{Observational results}
\label{obsresults}
The SEDs of all four systems have infrared excesses, indicating the presence of dust-rich disks. For two of the disks, Haro 6-13 and DO Tau, the {\it Herschel} B2A spectrum peaks at 63$\mu$m with a shape consistent with the crystalline water ice feature \citep[and see opacities in Figure \ref{wateropas}]{bertie+69}.  AA Tau displays no crystalline features, but the short wavelength end of the B2A spectrum continues up into the long wavelength shoulder of the 47$\mu$m amorphous ice feature.  The ISO spectrum confirms the existence of the amorphous 47$\mu$m feature. In contrast to these three objects, VW Cha shows no ice feature and is essentially smooth over its PACS spectrum.

For DO Tau, Haro 6-13, and VW Cha we calculated the equivalent width of the over the 63$\mu$m region, which captures potential emission from the crystalline ice feature.  The equivalent with is defined as:
\begin{equation}
 W_{\lambda}=\int_{\lambda_1}^{\lambda_2} \! \frac{F_{obs}-F_{continuum}}{F_{continuum}} \, \mathrm{d}\lambda
 \end{equation}
 \noindent In both cases, the continuum was determined by a least-squares polynomial fit to the 25 and 32$\mu$m region of the {\it Spitzer} IRS spectrum and the 105 and 140$\mu$m region of the {\it Herschel} PACS R1S spectrum. The limits of integration for the crystalline 63$\mu$m feature were taken to be 56 and 72.5$\mu$m and are indicated in Figure \ref{ew63}.  Polynomials of the fifth degree produced continuua that were most consistent with the model fits found in Section \ref{input}; however, differences in $W_{\lambda}$ calculated with lower degree polynomial continuua was not greater than 3$\sigma$. This figure also shows the polynomial continuum fits for the two disks in which we detect the 63$\mu$m feature, DO Tau and Haro 6-13, and the disk in which there is a non-detection of ice, VW Cha. For comparison, we include GQ Lup, the T Tauri disk in which we first detected the 63$\mu$m crystalline ice feature in \citet{mcclure+12}.   In Figure \ref{ew63} we list as well the 1$\sigma$ uncertainties for the equivalent widths, calculated from the individual error bars in the spectra over which they were taken. Spectral energy distributions of disks have intrinsic curvature in the 55 to 120$\mu$m region due to optical depth and temperature effects, leading to a systematic uncertainty in the location of the continuum underlying the feature.  We use the calculated value of the equivalent width for VW Cha, in which ice is not detected, to represent the this uncertainty.  This systematic uncertainty is less than three times the flux-based uncertainties.

\begin{deluxetable}{ccc}
\tabletypesize{\small}   
\tablewidth{0pt}
\tablecaption{Equivalent width of 63$\mu$m region}
\tablehead{\colhead{Star} & \colhead{$W_{63\mu m}$} & \colhead{$\sigma_{63\mu m}^a$} \\
                    \colhead{ }   & \colhead{($\mu$m)}  & \colhead{($\mu$m)} }
\startdata
VW Cha$^b$	&	0.5	& 0.2 	\\
GQ Lup  	&	1.6	& 0.2 	\\
DO Tau	&	2.1	& 0.3 	\\
Haro 6-13	&	3.3	& 0.5  	
\enddata
\tablecomments{
\\
$^a$ Statistical uncertainty calculated from the spectral flux uncertainties.
$^b$ Ice was not detected in VW Cha. This equivalent width represents the systematic uncertainty arising from the intrinsic curvature in disk continuua over the far-infrared spectral region.}
\label{ewtab}
\end{deluxetable}

\section{Analysis}
\label{allanalysis}

The goal of this study is to deduce the water ice distribution and abundance in these systems. In an optically thick disk, the relative strength (i.e. equivalent width) of any dust feature depends on the abundance of the emitting dust species, grain geometry and size distribution, degree of dust settling, disk temperature and density structures, and inclination of the disk along the line of sight \citep{furlan+09}.  It is therefore possible to use the observed 63$\mu$m and 47$\mu$m features to infer these properties by constructing a model disk structure and using radiative transfer to produce a synthetic spectrum to compare with the observed data.  

Our approach to the analysis of these objects is as follows.  First, we fit the SEDs of each system with an emergent spectrum from the \citet{dalessio+06} disk structure models using specific dust opacities (Section \ref{sedmods}).  This allows us to constrain the grain size distribution of the ice, the degree of dust settling, and disk temperature structure.  We can also use the radiative transfer calculation to determine where the emitting region is for the water ice (Section \ref{emit}).  To assess the effect to the overall ice distribution of photodesorption of ice mantles from grains, we run a separate model to compute the UV radiation field and combine the results with the density structure found from the D'Alessio models to determine the disk location where water ice has been completely photodesorbed (Section \ref{uvmodels}).
\subsection{Disk structure models}
\label{sedmods}
The detailed physics of the D'Alessio irradiated accretion disk models is described in \citet{dalessio+98,dalessio+99,dalessio+01} and \citet{dalessio+04,dalessio+06}.  For the benefit of the reader, we briefly highlight the main features of the code, and our updates to it, as it pertains to this study.  
\subsubsection{General description}
The \citet{dalessio+06} codes calculate self-consistently both the density and temperature structures of a disk heated by both stellar irradiation {\it and} viscous dissipation.  The mass accretion rate, $\Mdot$, is constant throughout the disk. Viscosity is parameterized through $\alpha$ \citep{shakura_sunyaev73}, which is also held constant. The disk consists of gas and dust, with the dust divided into zones to simulate spatial variation in its properties and composition.  

Vertically, there are two dust populations: the upper disk layer population has a smaller maximum grain size, $a_{max}$, while the midplane population has much larger $a_{max}$.  Settling from the upper layers to the midplane is parameterized through depletion of the upper layer population, $\epsilon=\xi/\xi_{standard}$, where the denominator is the sum of the mass fraction of the different components relative to gas and the numerator is the mass fraction in the small dust population.  The midplane dust abundance is correspondingly enhanced to account for material removed from the upper layers.  

\subsubsection{Updates}
Radially, we have updated the D'Alessio code to include multiple dust populations with discrete transitions at cut-off radii ($R_C$).  A more comprehensive exploration of the effects of radial zoning is left to future work; here we consider a two-zone model {\it only} in the event that a the standard D'Alessio single-zone model fails to reproduce simultaneously the PACS ice feature and the IRS slope. The two-zone model allows us to vary the dust properties in the upper or lower layer population interior or exterior to $R_C$.  In any zone, radial or vertical, the dust properties, such as the grain size distribution, abundance, and opacity can be changed individually for each of three main grain types: silicates, pure, solid carbon, and water ice.  The details of the dust are described in Section \ref{opas}.

At the inner edge of the disk, we implement a vertical dust sublimation wall with an atmosphere following the prescription of \citet{dalessio+04} with the two-layer curvature proposed by \citet{mcclure+13b}.  To recap briefly, there is a lower-layer, with height $h_{wall,1}$ and a rim $z$ coordinate of $z_{wall,1}$, and an upper layer of height $h_{wall,2}$ = $z_{wall,2}-z_{wall,1}$ to produce a simple box-function shape.  Each wall layer is characterized by its grain size distribution and sublimation temperature.  Beyond each layer's radius, $R_{wall}$ defined by those two properties, dust is present. We allow the dust properties of the wall layers to vary independently of those in the disk to simulate the effects of a radial gradient in the inner disk mineralogy. 

\subsubsection{Opacities}
\label{opas}
The dust opacities are calculated from optical constants using Mie theory under the assumption of segregated, spherical grains. The grain size distribution of each grain type is of the form $n(a)=n_0a^{p}$, where $a$ is the grain radius with limits of 0.005$\mu$m and $a_{max}$.  The abundance of each species is input as a mass fraction relative to gas.  For the silicate and graphite grains, we assume $p=-3.5$ \citep{mrn77}, and the optical constants of graphite are taken from \citet{dl84}.  Silicates are divided into amorphous and crystalline versions of two stoichiometries: olivines ($Mg_{2-2x}Fe_{2x}SiO_4$) and pyroxenes ($Mg_{1-x}Fe_{x}SiO_3$), where $x=Fe/(Fe+Mg)$ indicates the iron content.  Opacities for the amorphous olivine and pyroxene are computed with optical constants from \citet{dorschner+95} that have $x=0.5$ for olivine and range from $0.05$ to $0.6$ for pyroxene.  The opacities for crystalline olivine (forsterite) and pyroxene (enstatite) are taken directly from those calculated by \citet{sargent+09b}.  The mass fractions of silicates and graphite are fixed at 0.004 and 0.0025, respectively.

For the water ice grains, we used the optical constants compiled by \citet{warren_brandt_08}.  These constants are, however, specifically designed for 266 K and atmospheric pressures, and in the lower pressures of protoplanetary disks, water ice sublimates at lower temperatures.  For this reason, we created hybrid optical constants using the \citet{curtis+05} optical contants for crystalline (146 K) and amorphous (106 K) ice over the $\sim$15 - 200$\mu$m range and the \citet{warren_brandt_08} optical constants outside of that wavelength range.  The details of the opacity slicing are shown in Figure \ref{wateropas}.  The amorphous ice is characterized by a single peak at $\sim$47$\mu$m with a broad wing towards 100$\mu$m, while the crystalline ice has two peaks at 43$\mu$m and 63$\mu$m.  As shown in \citet{curtis+05}, these peaks are shifted to shorter wavelengths with decreasing temperature.  

\subsubsection{Models for individual disks}
\label{input}
For each object, we varied $\epsilon$, $\alpha$, the maximum grain size in the upper layers, $a_{max,s}$, and the ice mass fraction relative to the gas to determine the best-fitting range to the SED and the observed ice feature.  Other essential inputs are the stellar, accretion, and disk parameters, i.e. $T_{eff}$, $R_*$, $M_*$, $\Mdot$, $A_V$, $i$, and $R_d$.  These were taken from the literature and held fixed.  We describe below our choices for the fixed properties and describe briefly each disk's specific modeling challenges. The final properties, fixed and fitted, are listed in Table \ref{modelfits}.   

{\it VW Cha}: This is a triple system composed of a primary separated by 0\farcs65 - 0\farcs7 from its secondary, which is itself binary with a 0\farcs1 projected separation \citep{brandeker+01}.  Spectroastrometry indicates that the primary is a K5 star with an equivalent width of H$\alpha$, $W_{H\alpha}$, of 79 \AA, while the emission from the secondary and tertiary has a combined K7 spectral type (SpT) and $W_{H\alpha}$ of 4.8 \AA.  Additionally, the primary is a factor of 4 brighter at {\it K}-band than the total emission from the secondary and tertiary components \citep{ghez+97}.  Together this suggests that emission from an accreting circumprimary disk dominates the SED.  Consequently, we do not correct for the contribution of the combined secondary and tertiary stellar components to the total emission.  For the stellar and accretion parameters, we take the values for the primary given by \citet{hartmann+98} and the $A_V$ found by \citet{gm03}.

\begin{figure}
\centering{\includegraphics[angle=0, scale=0.7]{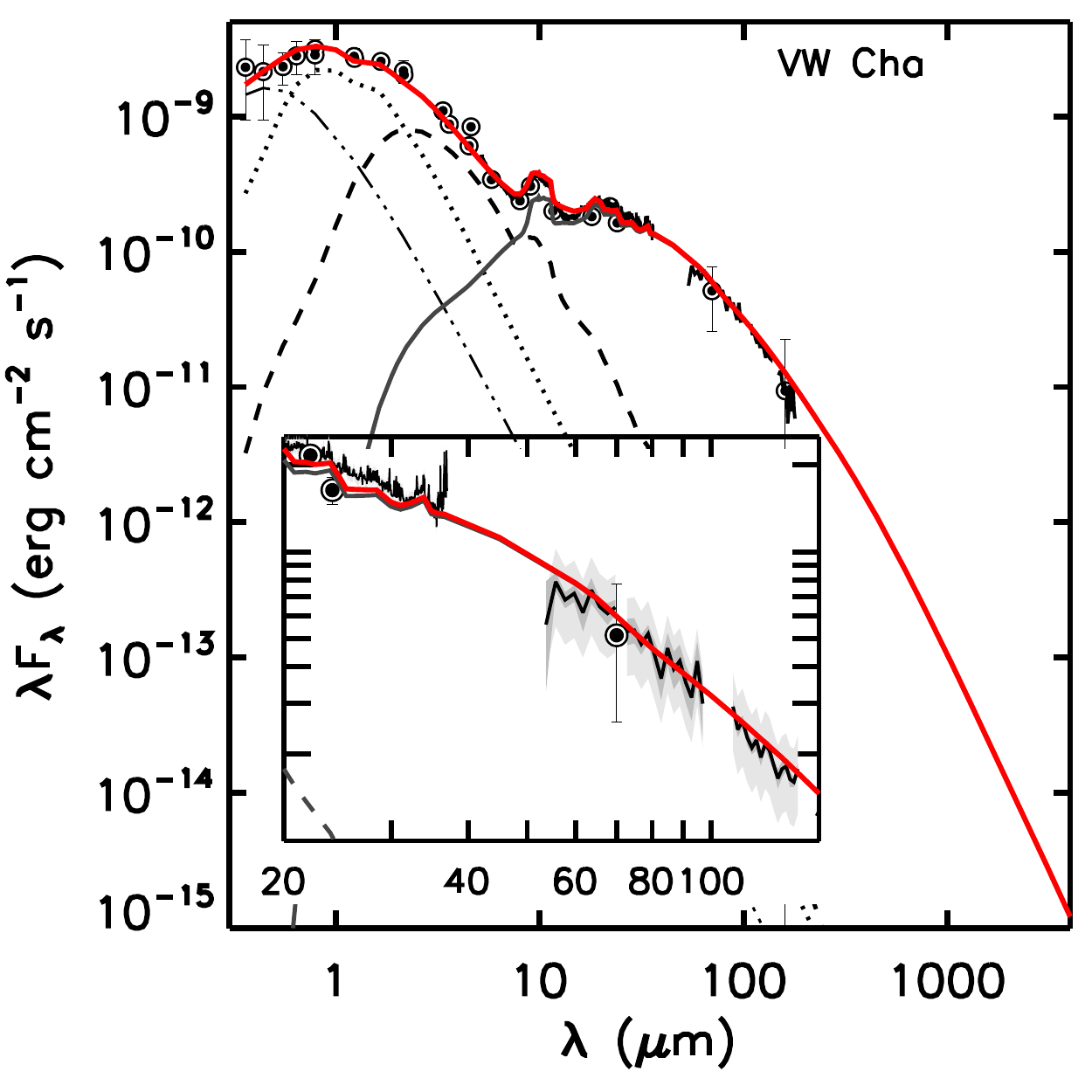}}
\caption{Best-fitting model fit to VW Cha. Spectral and photometric observations are plotted in thick black; photometric errors are 3$\sigma$. Light gray band around PACS spectrum represents the absolute flux calibration uncertainty of 30\%.  The dark gray band represents the point-to-point 3$\sigma$ uncertainty in the spectrum, i.e. the uncertainty in the shape. 
Total model fit (red) includes the following components: accretion excess (dash-dotted line), stellar photosphere (dotted line), curved wall (dashed line), and disk (solid line). Best-fitting parameters are given in Table \ref{modelfits}. The best fitting model does include ice, but no feature is visible, consistent with the feature arising in the outer disk.  \label{vwcha1}}
\end{figure}

The inferred circumprimary disk has not been spatially resolved.  Uncertainty in $R_{*}$ and $v$sin$i$ makes definitive estimates of the inclination angle difficult; after testing several values, we find that models between 20 and 45\degr produce a good fit to the data.  We assume 45\degr for the remainder of the paper, similar to the inclinations of Haro 6-13 and DO Tau.  The upper limit to the outer radius, $R_d$, predicted from disk-binary interaction theory is 0.4 times the AB separation, assuming a circular orbit \citep{al94}. With a de-projected separation of 147AU, $R_d$=59AU. Smaller values of $R_d$, e.g. 10 to 30AU, fit better suggesting some eccentricity in the system.  This may be an effect of the tertiary component.  The appearance of the 10 and 20$\mu$m silicate complexes is similar to disks with strong crystalline silicate indices in \citet{watson+09} and the IRS region is well-fit with forsterite in our model, consistent with a smaller, hotter disk.  The best-fitting model is given in Figure \ref{vwcha1}.

\begin{figure}
\centering{\includegraphics[angle=0, scale=0.7]{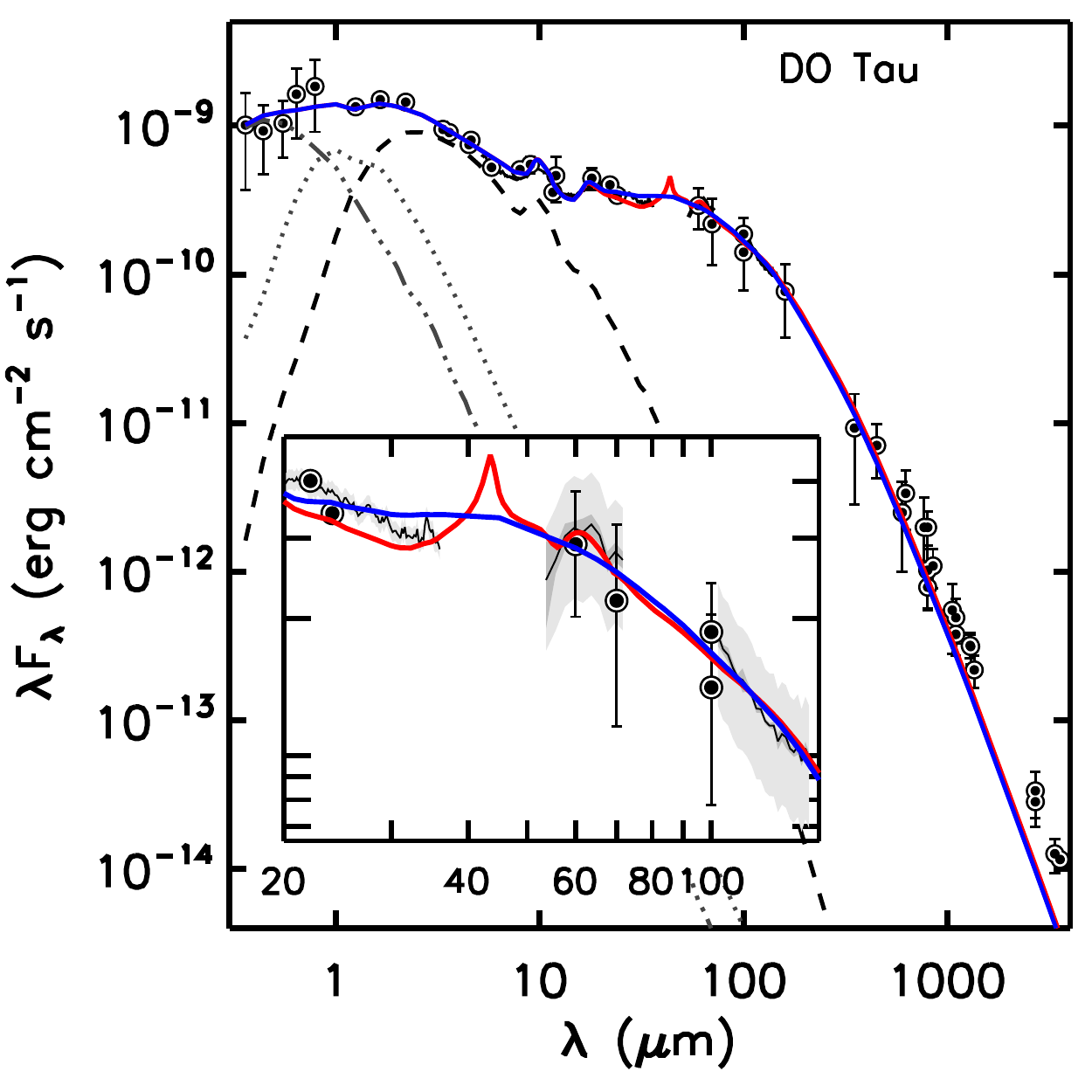}}
\caption{Best-fitting model fit to DO Tau. Plotting styles are the same as in Figure \ref{vwcha1}. The red curve is the model fit with ice and the blue curve is the model fit with only silicates and graphite. \label{dotau1}}
\end{figure}

{\it DO Tau}: This is a marginally resolved system.  We use the radius and inclination found by \citet{koerner+95}.  Taking the spectral type given by \citet{kh95}, we use an IRTF SpeX spectrum (McClure et al., in prep.) to compute the veiling, extinction, luminosity, and mass accretion rate as described in \citet{mcclure+13a}, using the weak-line T Tauri star LkCa 14 as a photospheric template.  DO Tau displays signs of radial variation in its gas and dust content; the millimeter dust continuum has a smaller radial extent than the gas line emission or scattered light from small dust grains \citep{koerner+95,kitamura+02,itoh+08}.  The scattered light images show an asymmetric arc at $\sim$250AU \citep{itoh+08} that suggests dynamical effects in the outer disk. Radiative effects may play a role as well; DO Tau has a strong UV radiation field, with $L_{FUV}$=0.0325$L_{\odot}$ \citep{yang+12}. 

\begin{figure}
\centering{\includegraphics[angle=0, scale=0.7]{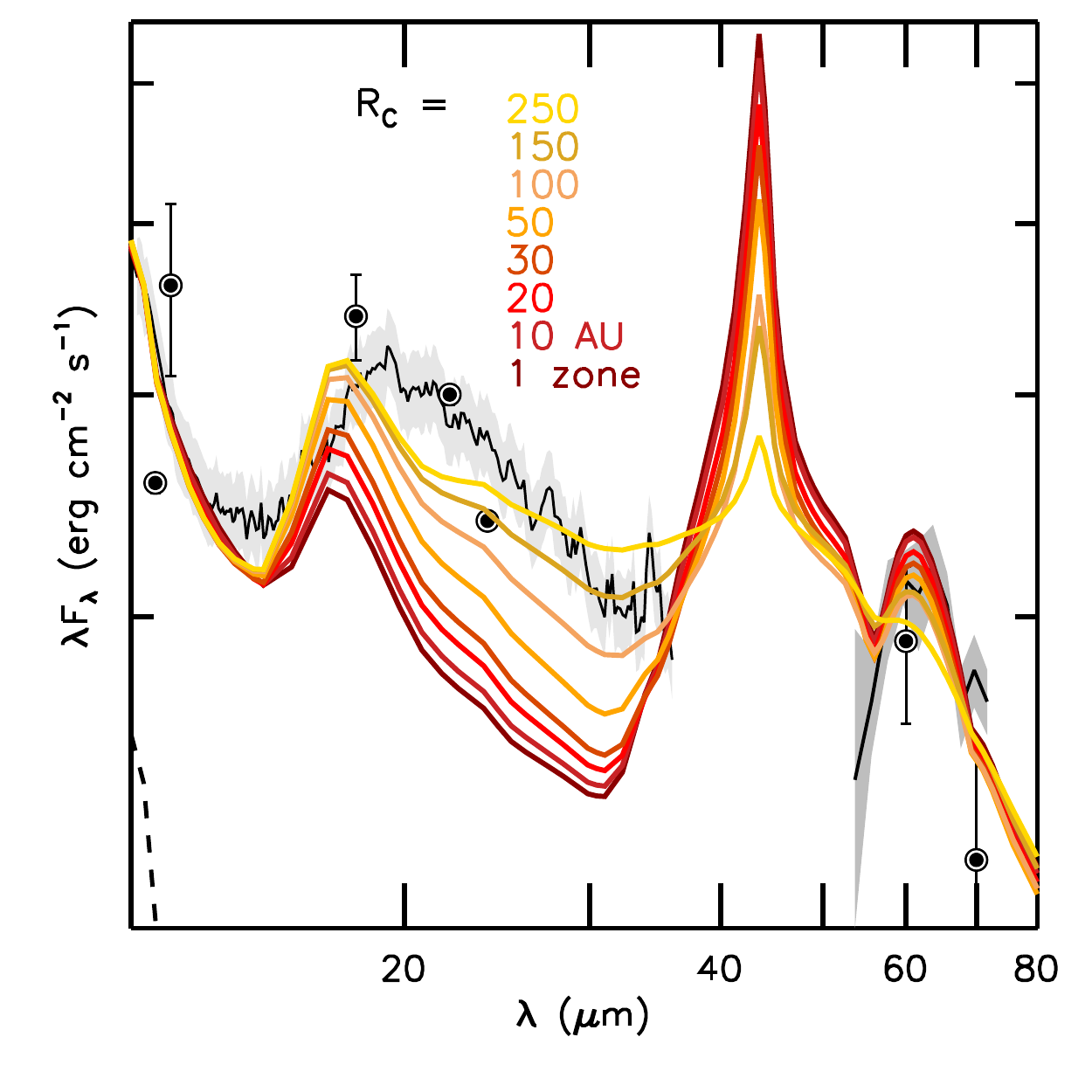}}
\caption{Effect on the shape of DO Tau's SED of lowering to 0.00001 the water ice mass fraction in the disk upper layers interior to some radius, $R_C$. Exterior to $R_C$, ice is present in the upper layers with a mass fraction of 0.002.  Ice in the midplane layer has the same mass fraction in both the R$<$R$_C$ and R$>$R$_C$ zones.  \label{dotau_rcs}} \end{figure}

The best fit to DO Tau is shown in Figure \ref{dotau1}.  In fitting the SED of DO Tau, we were able to fit simultaneously the shape of the IRS spectrum and the PACS spectrum with a radially uniform disk model, but not the strength of the 63$\mu$m ice feature.  Models that fit the ice feature underfit the 20-30$\mu$m observed flux.  To mimic the effect of the dust arc in the outer disk observed by \citet{itoh+08}, we used the modified version of our code to simulate an outer ring of water ice. Although the model that fits best the strength of the ice feature still produces too little flux from 20 to 30$\mu$m, this effect is weaker for the 2-zone model fit than for the radially uniform case (see Figure \ref{dotau_rcs}).

{\it Haro 6-13}:  We take the spectral type found by \citet{wh04} and extinction correct the optical and NIR photometry to match those of LkCa 14.  This yields an $A_V$ of 7.6 magnitudes; however, including the four veiling estimates given by \citet{wh04} and \citet{fe99} in the extinction calculation using the method described in \citet{mcclure+13a} yields $A_V$=6 magnitudes.  Due to discrepancies between the dwarf and WTTS colors and uncertainty in the $A_V$ of LkCa14, we adopt an $A_V$ of 7.0 magnitudes.  The stellar luminosity and radius then follow; the mass is calculated from the \citet{siess+00} tracks, and the mass accretion rate calculated from Br$\gamma$ using the calibration of \citet{muzerolle+98} and the Br$\gamma$ luminosity found by \citet{gl96}. 

\begin{figure}
\centering{\includegraphics[angle=0, scale=0.7]{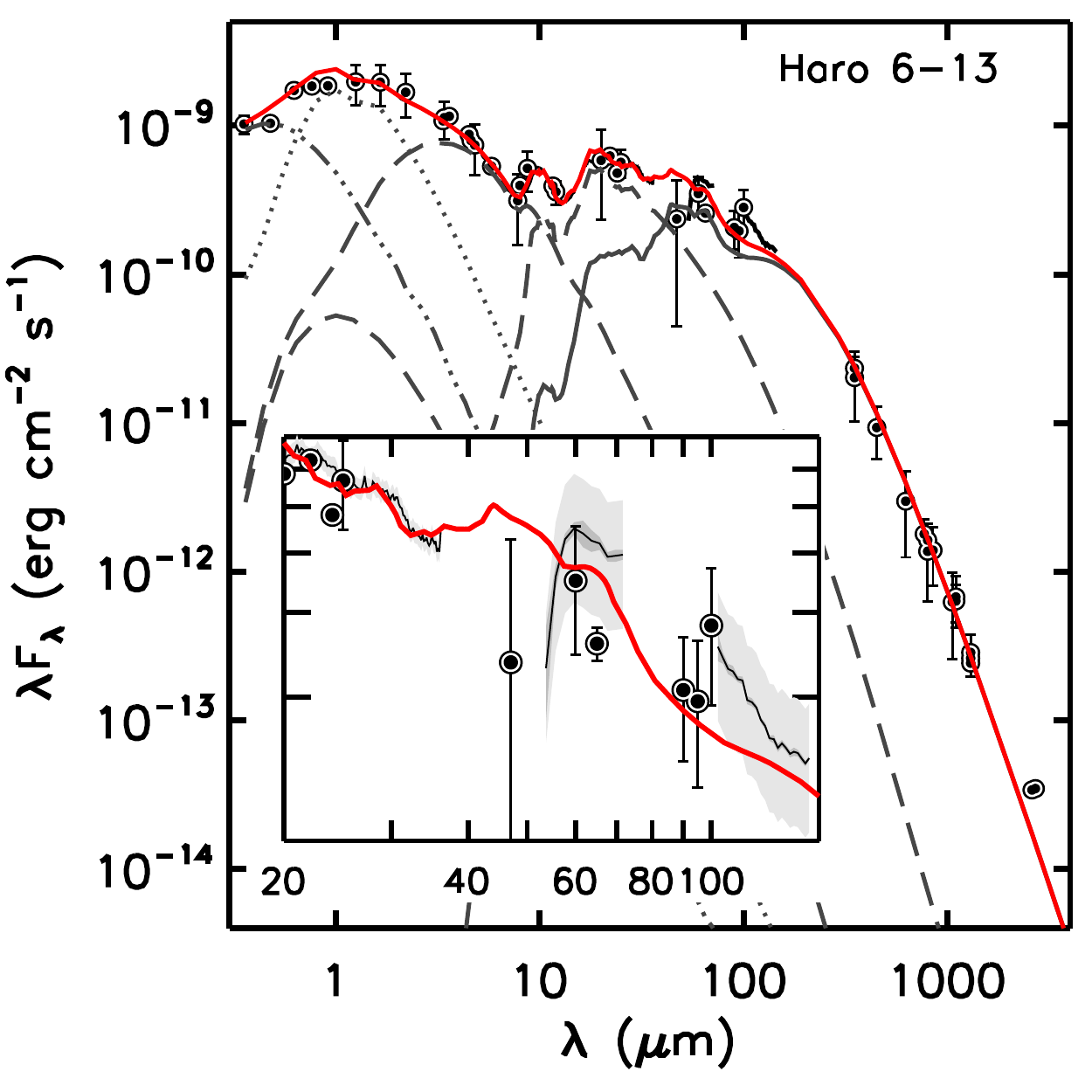}}
\caption{Best-fitting model fit to Haro 6-13. There are two wall components; one for the inner wall and one for the outer wall.  Both are plotted as dashed lines.  Other plotting styles are the same as in Figure \ref{vwcha1}.  \label{haro6131}}
\end{figure}

The spectral slopes of Haro 6-13 between 6, 13, and 31$\mu$m \citep{watson+09} are consistent with those of pre-transitional disks in Taurus and Ophiuchus \citep{mcclure+10}, suggesting that this disk may have a gap \citep{espaillat+14}. Attempts to fit Haro 6-13 with a radially uniform disk were unsuccessful; a gap of $\sim$6AU was required to reproduce the sharp increase in flux between 13 and 31$\mu$m.  As with DO Tau, we were able to reproduce well the overall SED, but not in conjunction with the 63$\mu$m ice feature.  It appears that either the absolute flux of the PACS spectrum is too high or the photometry is too low, although the 30\% absolute flux uncertainty and 3$\sigma$ error bars of the far-infrared photometry overlap. The SED may also dip near 40$\mu$m, which we do not attempt to reproduce.  The models demonstrate that there is ice detected in this disk, but the poor fit over the far-infrared suggests additional structure that we cannot take into account with our current models. The best-fitting model is displayed in Figure \ref{haro6131}.  In Table \ref{modelfits}, we include the parameters of the outer disk wall as a footnote, for the sake of uniform formatting.

\begin{figure}
\centering{\includegraphics[angle=0, scale=0.7]{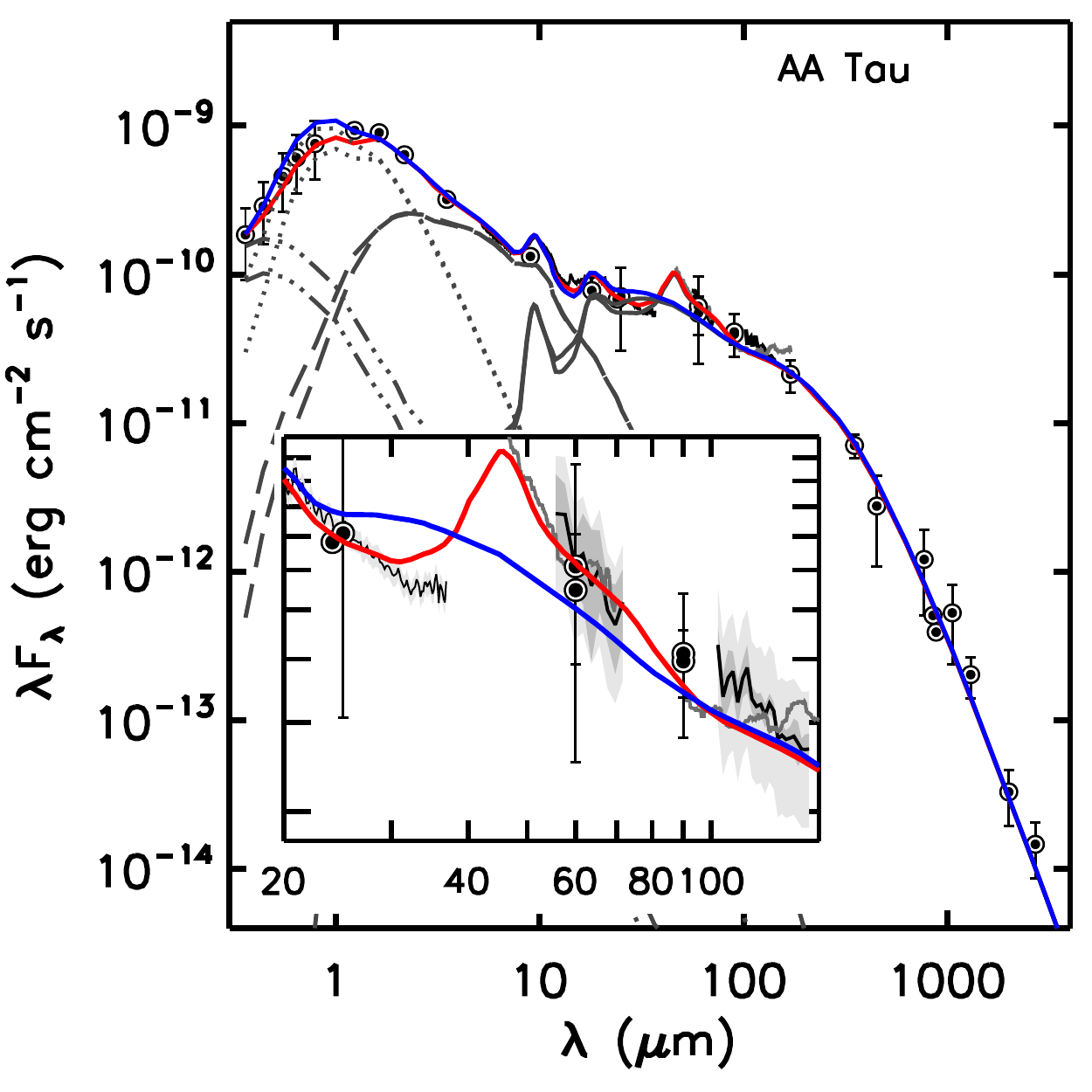}}
\caption{Best-fitting model fit to AA Tau.  Plotting styles are the same as in Figure \ref{vwcha1}. The red curve is the model fit including ice and the blue curve is the model fit with only silicates and graphite. Additional dark gray spectrum from 43$\mu$m to 160$\mu$m is the archive ISO LWS spectrum, trimmed and scaled.  \label{aatau1}}
\end{figure}

{\it AA Tau}: This is a spatially resolved system at a higher inclination than our other targets \citep{kitamura+02}.  We assume the stellar luminosity and effective temperature given by \citet{hartmann+98}, recompute the mass using the \citet{siess+00} tracks, and calculate the mass accretion rate from the luminosity of the Br$\gamma$ line via the relation given by \citet{muzerolle+98}, using the equivalent width of Br $\gamma$ given by \citet{fischer+11} and the maximum K band flux given by \citet{eisner+07}.  We take $A_V$ from \citet{fischer+11} and the inclination and outer disk radius from \citet{cox+13}.  We note that the radius found by \citet{cox+13} from coronographic images is smaller than that found from submillimeter observations \citep{aw07}; the former authors attribute the difference to contamination from a background source in the submillimeter image.

There is still a small discrepancy between the best-fitting model and data at the end of the IRS range, but this difference may arise due to a number of convergent effects that are difficult to account for.  In particular, the location of the peak of the 47$\mu$m feature between the IRS and ISO spectra, the difference in epochs between these spectra, and potential radial or vertical variation in the ice grain size could all affect the shape of this feature relative to the continuum. Despite this issue, the model fit to AA Tau is the best out of the three disks with ice detections and, overall, fits well both the SED and feature strength. The best-fitting model is given in Figure \ref{aatau1}.


\subsection{Ice emitting region}
\label{emit}

Even though our spectral features are spatially unresolved, and therefore in theory sample the whole disk, in practice certain regions of the disk contribute more than others to the features we detect.  For example, if the optical depth along the line of sight exceeds $\sim$1, regions at higher optical depths (e.g. on the opposite side of the midplane) are invisible to us. Likewise, regions that are very cold or hot will emit at longer or shorter wavelengths, respectively, than our PACS spectra.  Using the opacity, density, and temperature of our best-fitting disk structure models, we can determine the contribution of each point in the disk to the observed intensity at a given wavelength.  This contribution is quantified by the integrand of the solution to the transfer equation for the emergent intensity:

\begin{equation}
I_{\nu}(0)=\int_0^{+\infty} \! S_{\nu} e^{-\tau_{zp}} \, \mathrm{d}\tau_{zp}
\end{equation} 

\noindent where $\tau_{zp}$ is the optical depth calculated along the line of sight towards the observer, taking into account the inclination of the disk.  Since we are interested in the dust, the source function, $S_{\nu}$, is the Planck function.  We rewrite the integrand in terms of the coordinate along the ray in the line of sight, $z_{p}$ and define a `contribution function' of $z_{p}$ as:

\begin{equation}
C(\nu,z_p)=B_{\nu}(T) e^{-\tau_{zp}} \frac{\mathrm{d}\tau_{zp}}{\mathrm{d}z_p}\mathrm{d}z_p
\end{equation}

\noindent where the differential expression is simply equal to the opacity at $z_{p}$.  By definition this function will be negligible in regions of the disk where $\tau_{zp}$ is either very large or relatively constant.  It also will be smaller in cooler regions of the disk.  As $\tau_{zp}$ is dependent upon the inclination angle along the line of sight, the contribution function will be asymmetric between the near and far sides of the disk.  Water ice contributes to both the flux on either side of the feature and the feature itself. Therefore to identify the emitting regions probed specifically by the 47 and 63$\mu$m features, one can subtract the contribution to the emission at 72$\mu$m from the contribution to the emission at the peak wavelength of each ice feature for each spatial grid point in the disk structure model. 

\begin{figure*}
\centering{\includegraphics[angle=0, scale=0.71]{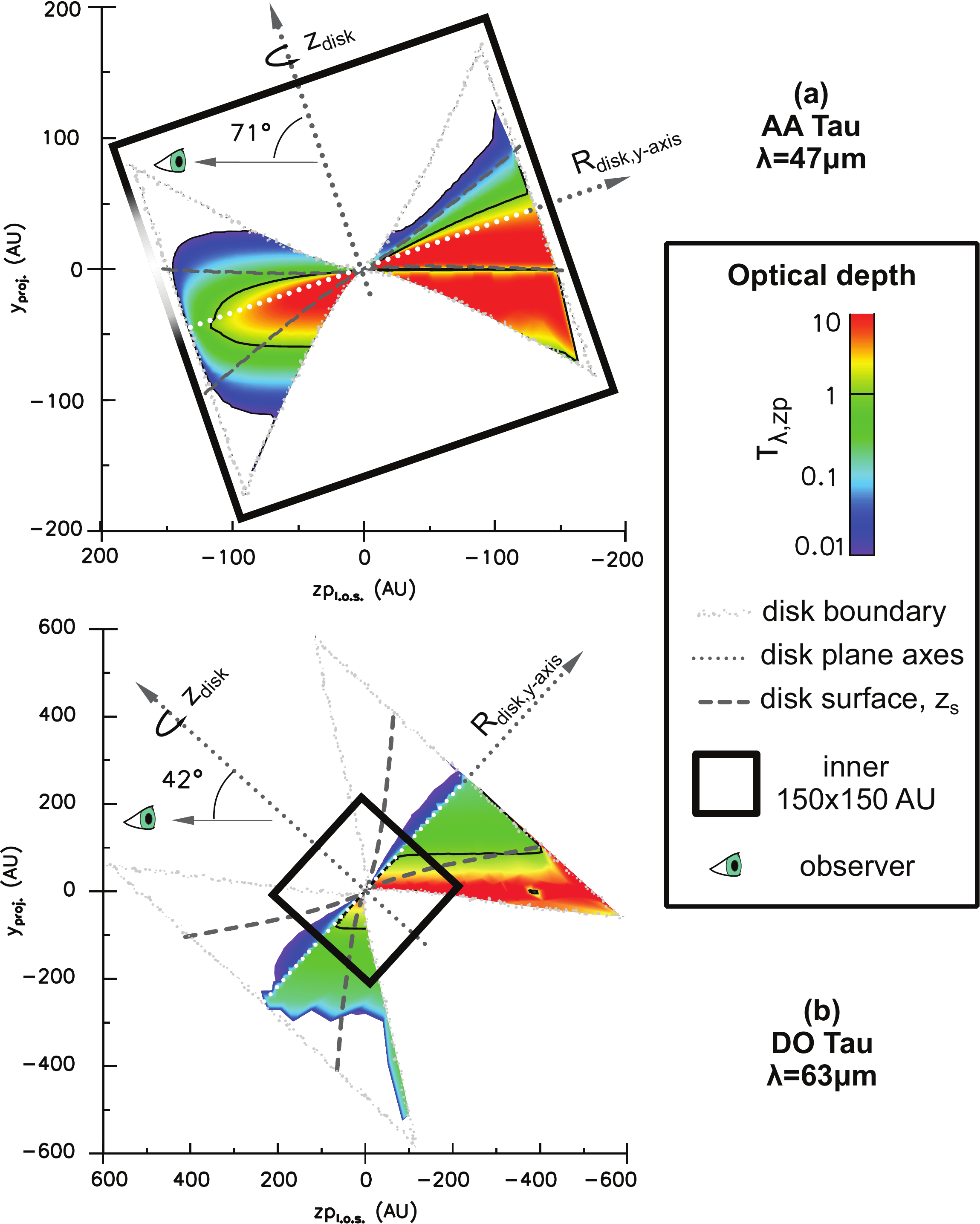}}
\caption{Comparison of the optical depth along the line-of-sight (zp) in the plane of the sky for AA Tau (top) and DO Tau (bottom) for the 47$\mu$m amorphous ice and 63$\mu$m crystalline ice features, respectively.  The isocontour of $\tau_{\lambda,zp}$=1 is indicated.  The optical depth structures are overlaid with the disk geometry, including the maximum extent of the disk models in our calculations, and the mid plane and disk rotation axis.  The disk surfaces, where $\tau$=1 to stellar radiation, are also displayed. Linestyles and symbols are defined in the legend. \label{tauzps}} 
\end{figure*}

\begin{figure*}
\centering{\includegraphics[angle=0, scale=0.8]{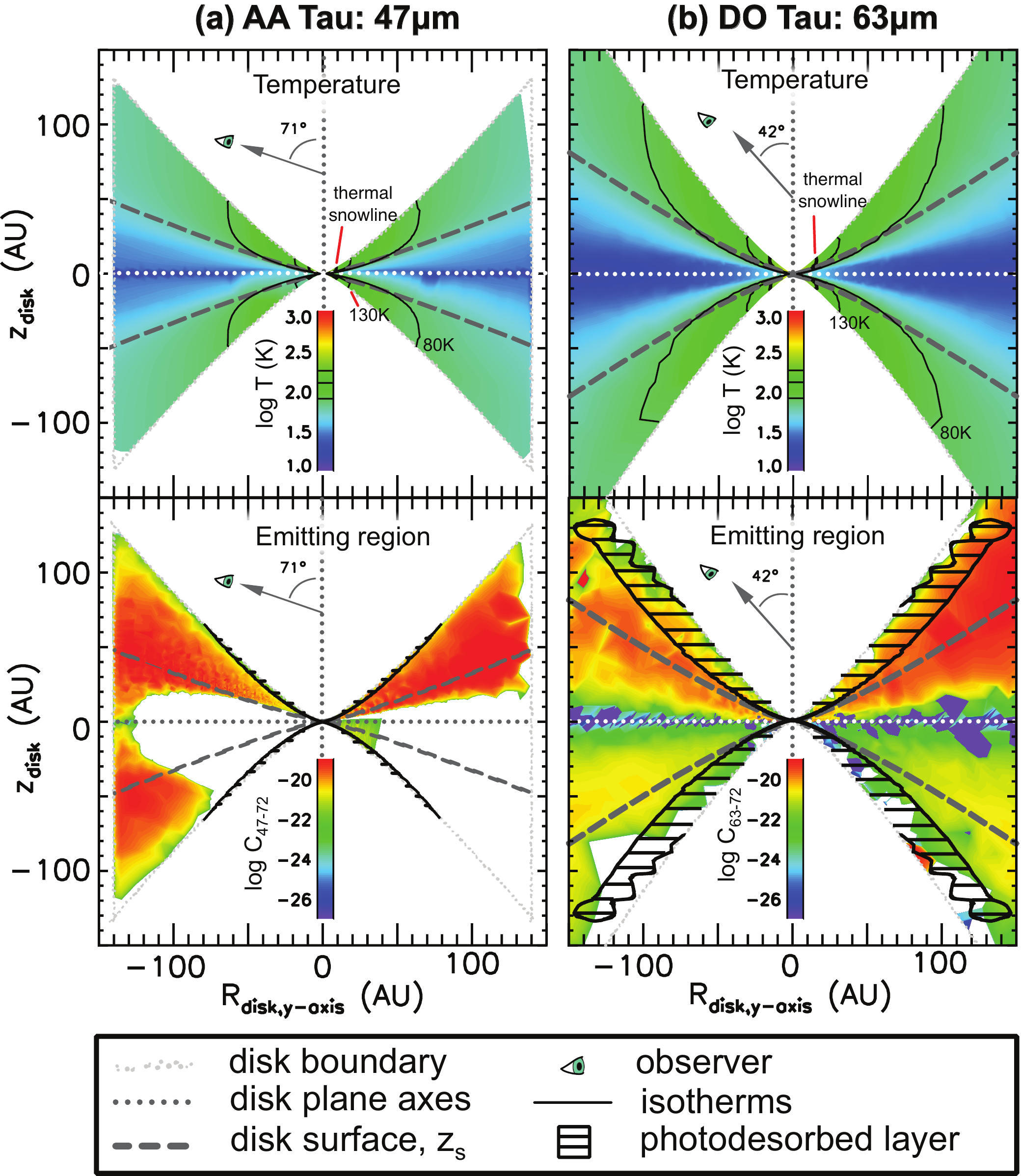}}
\caption{Comparison of the temperature structure and emitting area contributing to the ice features for AA Tau (left) and DO Tau (right).  The emitting region is calculated by subtracting the contribution to the total intensity of each point at 72$\mu$m from that at either 47 or 63$\mu$m, as described in the text. Note that the red regions contribute most to the flux while purple or white regions contribute the least. The regions of the disks above the assumed equilibrium ice crystallization temperature, 130K, are indicated, along with the regions below 80K, where high energy particles will re-amorphize crystalline ice after 0.1 Myr.  The photodesorption layer, where ice is completely removed from the grains, is indicated. For AA Tau, the latter is a barely visible thin skin in the upper layers of the disk.  \label{cont_funct}} \end{figure*}

Of the three new ice detections, Haro 6-13 has the poorest model fit, potential radial disk structure, and no ancillary UV data (for use in \ref{uvmodels}).  We exclude it from further analysis and concentrate our efforts on AA Tau and DO Tau. For both of these disks, we compute the optical depth along the line of sight, the temperature structure, and the final, differenced contribution function using the uniform disk model that best-fit the ice feature. We note that in the case of DO Tau, the model used is not the overall best-fitting model, but a uniform upper layer ice abundance is required for this comparison. Figures \ref{tauzps} and \ref{cont_funct} show the results.  

\subsection{UV radiation field models}
\label{uvmodels}

The weakness of cold water vapor detections from {\it Herschel} \citep{hogerheijde+11} may indicate that most of the grains exposed to the UV radiation field are not icy.  Rather it is likely that the icy grains lie just below the UV-exposed layer. To investigate the effect of ultraviolet radiation on the distribution of water in these disks, we wish to compare the emitting region identified by the contribution function (Section \ref{emit}) with the disk region in which water ice would be completely photodesorbed from grains.  Using the \citet{bethellbergin09} code, we calculate the UV radiation field at each point in the disk, taking our best-fitting disk structure as input and assuming an input UV spectrum of TW Hya, scaled to the luminosities of AA Tau and DO Tau given by \citet{yang+12}.  The location at which photodesorption of ice from a grain overwhelms H$_2$O freeze-out on a grain in molecular clouds can be modeled as $A_V\propto ln(G_0Y/n)$, where where $G_0$ is the mean interstellar flux in units of Habing (1 Habing = $1.6\times10^{-3} erg/cm^2/s$), $Y$ is the photodesorption yield, and $n$ is the gas density \citep{hollenbach+09,boogert+13}.  If we assume that this relation holds for the upper layers of protoplanetary disks as well, then the surface in the disk where ice no longer exists should be the set of points (R, z) that satisfy:


\begin{equation}
G_{0,disk}(R,z)=\frac{n_{disk}(R,z)G_{0,cloud}}{n_{cloud}}
\end{equation}

\noindent We calculate $G_{0,disk}(R,z)$ from the \citet{bethellbergin09} flux; combining these $G_{0,disk}(R,z)$, with the assumption of 1 and 10$^3$cm$^{-3}$ \citep{hollenbach+09} for the values of $G_{0,cloud}$  and $n_{cloud}$, respectively, we derive the photodesorption surface for both disks (see Figure \ref{tauzps}, bottom right panel).  

\section{Results of analysis}
\label{results}

The details of how we fit each source are given above (Section \ref{input}).  In general good fits to the data were obtained with radially uniform disks for VW Cha and AA Tau, while radial variations were required to fit Haro 6-13 and DO Tau.  The modifications still do not produce ideal fits. Despite this, we can extract useful information about the the ice distribution from the sample as a whole.

\subsection{Dust settling vs ice abundance}
To fit the overall far-infrared fluxes and slopes of their SEDs, these disks required a local enhancement of the dust/gas ratio at the midplane of a factor of $\sim$10 (corresponding to $\epsilon\sim0.1$) and midplane grain growth to millimeter size pebbles in order to fit the SED submillimeter slopes.  However, to fit the absolute flux of the {\it Spitzer} and {\it Herschel} spectra of the two disks in which the crystalline 63$\mu$m feature was detected, the upper layers required less depletion than the mean depletion value for disks in Taurus and Ophiuchus \citep[][$\epsilon$=0.01-0.001]{mcclure+10}, meaning these disks are less settled.  This result is consistent with the suggestion by \citet{hogerheijde+11} that water ice has typically settled out of the disk upper layers.  The fact that AA Tau, the disk with the amorphous 47$\mu$m detection, was best fit by an average degree of dust depletion does not contradict these results; the strength of the 47$\mu$m feature increases with higher inclination for a fixed degree of dust settling, and AA Tau is inclined at 71$\degr$, $\sim$30$\degr$ more than the other targets.

All three of the disks with ice detections were fit best by ice mass fractions of 0.002 based on the peak-to-continuum ratio of the feature, which increased with total ice abundance. Although the strength of this feature also depends on the degree of dust settling, we can constrain the latter from the overall flux and slope from $\sim$31 to 100$\mu$m.  This mass fraction of ice is only 36\% of the mass fraction predicted to be in the disk \citep{pollack+94}, which may suggest that icy-coated grains do preferentially grow and settle to the midplane better than bare grains.  Despite the non-detection of an ice feature, VW Cha could be equally well-fit by either an iceless disk or an disk with the same ice mass fraction as the other disks.  Significantly, its disk appeared to be truncated close to 7AU, which we assume to be a result of its secondary and tertiary companions and viscous evolution.  In \S\ref{rdice}, we will appeal to this result as a model-independent verification that the 63$\mu$m crystalline ice feature itself cannot originate primarily from the inner disk.  
 
 \subsection{Radial distribution of ice}
\label{rdice}

In Section \ref{emit} we calculated the contribution of each point in the disk to the 47$\mu$m amorphous ice feature in AA Tau and the 63$\mu$m crystalline feature in DO Tau, relative to the emission at 72$\mu$m for each disk.  Comparing the results for both disks (Fig. \ref{cont_funct}), we see that the amorphous ice feature for AA Tau samples exclusively the upper layers of the whole disk; the midplane remains optically thick over the entire radial range.  In contrast, although the primary contribution of the crystalline 63$\mu$m ice feature in DO Tau is still from the upper layers, it is concentrated below the disk surface and includes faint emission from $z_{disk} < 20$AU, closer to the midplane in the inner 120 AU of the disk.  This may simply be an effect of inclination, as AA Tau is more inclined than DO Tau by 30$\degr$. However, these tests do confirm that the detected ice in both disks is trans-Neptunian ($R>$30AU), probing the region from which a proto-Kuiper belt would be expected to form.

As an independent test of this result, we plotted the calculated equivalent widths of the 63$\mu$m crystalline ice features (see Section \ref{obsresults}) against the disk sizes. In general for optically thick disks the strength of a feature depends not only on the dust properties but also on the disk temperature structure and geometry in the regions where the continuum and feature arise. However, the three disks with crystalline features (DO Tau, Haro 6-13, and GQ Lup) and our control system, VW Cha, have roughly similar central stars, inclinations, and disk parameters, simplifying the comparison. Barring any differences in dust properties between these disks, the equivalent widths should track the outer radii of the disks if, for example, the feature emission were dominated by the outer disk due to its increased solid angle or larger fraction of the total disk mass.  

\begin{figure}
\centering{\includegraphics[angle=0, scale=1.0]{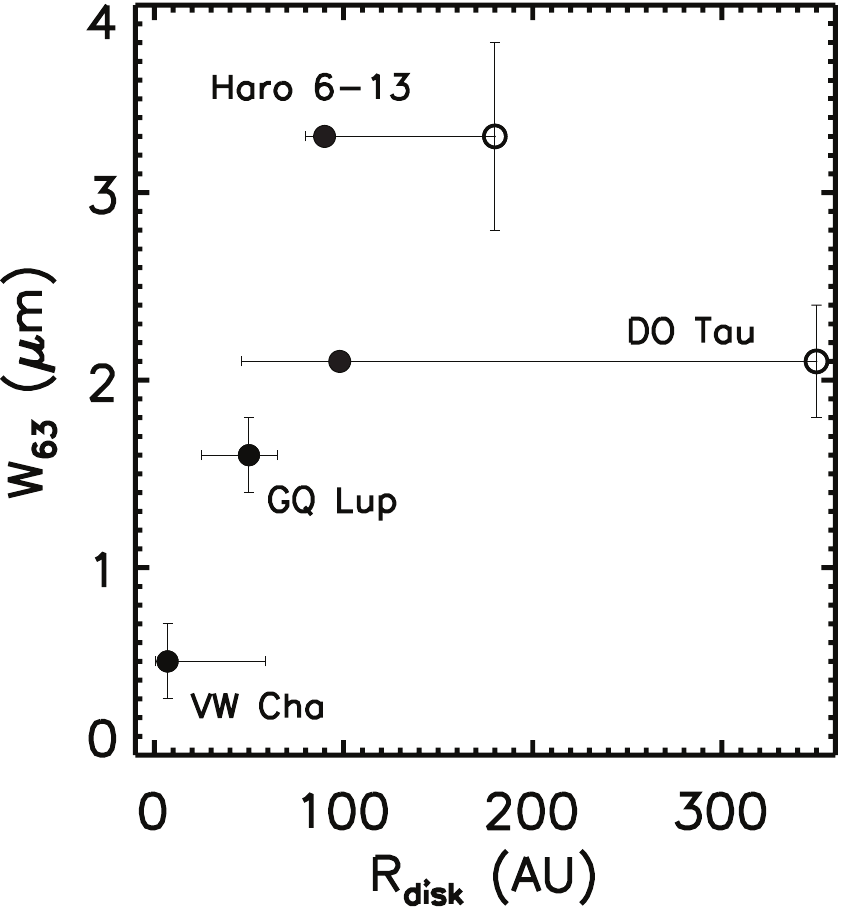}}
\caption{The equivalent width, $W_{\lambda}$, of the 63$\mu$m feature versus the outer disk radii for the two disks in which the 63$\mu$m feature is detected, the disk in which the feature is not detected (VW Cha), and our previously published feature detection in GQ Lup \citep{mcclure+12}. The disks of VW Cha and GQ Lup have hard outer limits imposed by the presence of their companions of 0.4 times their de-projected separations of 147 and 163AU, respectively, while the adopted value comes from the model fits to the SED.  For DO Tau and Haro 6-13, the open circle symbols indicate the radius derived for their CO disks (and presumably smaller, entrained grains), while the solid circles are for the millimeter grain disks. The equivalent width clearly correlates with the disk radius, suggesting the ice emission region is beyond 30AU. \label{icerad}}
\end{figure}

These outer radii are quasi-independent of model assumptions. For example, although their adopted disk radii of 7AU and 50AU come from the best-fitting model of their SEDs, the disks of VW Cha and GQ Lup have hard outer limits imposed by the presence of their companions of 59 and 65AU, or 0.4 times their de-projected separations of 147 and 163AU, respectively \citep[assuming circular orbits][]{al94}.  The dust and gas in DO Tau and Haro 6-13 have been imaged at submillimeter wavelengths, which provides a range of radii as the dust emission tracks millimeter grains while the gas emission tracks submicron grains that are coupled to the gas.  In Figure \ref{icerad} we plot $W_{63}$ against the disk radii.  The equivalent width shows a clear increase from a $\sim$0.5$\mu$m non-detection in VW Cha, $R_{disk}$=7AU, to 2.1$\mu$m for DO Tau, $R_{disk}$=350AU.  Our test suggests that the disks with larger outer radii have stronger ice features, likely due to a larger total ice mass.

\subsection{Vertical distribution of ice}
\label{uvresults}

Having determined that the 47 and 63$\mu$m ice features primarily sample the outer disk, we want to know where this ice is located vertically.  That these disks are less settled suggests that the water grains may be exposed to photodesorbing UV radiation.  On the other hand, ultraviolet radiation is extinguished at smaller penetration depths in the disk than visible light, and less settling means more optically thick disk upper layers, so the penetration depth should be even smaller than in more settled disks.  For AA Tau, it is clear that photodesorption does not significantly effect the ice distribution at large radii: in Figure \ref{cont_funct} the complete photodesorption surface is a thin skin of at most 5AU in vertical depth with a radial range out to 80AU.  This region is located well above that contributing to the 47$\mu$m ice emission feature. 

However, the $L_{FUV}$ of DO Tau is a factor of 16 greater than that of AA Tau \citep{yang+12}.  As demonstrated in the bottom right panel of Figure \ref{cont_funct}, this amount of UV radiation is enough to photodesorb ice out to 140AU in the disk, with a vertical depth of $\sim$20-30AU.  In fact, the photodesorbed region includes some portions of the disk that contribute to the 63$\mu$m ice feature and to the continuum between 20 and 30$\mu$m \citep{dalessio+06}.  These areas should be ice free, and although the majority of the 63$\mu$m contribution comes from below the photodesorption layer the abundance of ice in the lower disk layers may also be less than if it were below a region unaffected by photodesorption.  The combination of turbulence and photodesorption may reduce the amount of midplane water ice by moving midplane ice to the disk surface, where it is desorbed and reacts with other molecules \citep{furuya+13}.  Over time this cycle could deplete ice from the solid phase at given radius for all disk heights, as shown by \citet{furuya+13} for the inner 30AU of a disk.  Future spatially resolved studies probing the gas phase abundance of water and other molecules are necessary to confirm or refute this scenario.
 
\section{Discussion}
\label{discussion}

\subsection{Is crystalline ice a signature of planetesimal collisions?} 

To place these results in context, we can compare the inferred properties of our disk water ice detections with those of other disks at shorter wavelengths and solar system comets. Ice has been detected in absorption at 3$\mu$m in protostars and edge-on systems by \citet{pontoppidan+05}, \citet{terada+07}, \citet{honda+09}, \citet{terada_tokunaga12}, and \citet{aikawa+12}.  These authors are able to fit the 3$\mu$m feature best with grain sizes of 0.5 to 1$\mu$m and an abundance relative to hydrogen of 9$\times$10$^{-4}$ \citep{pontoppidan+05}, similar to the mass fraction of 2$\times$10$^{-3}$ found in this work.  In only two of these detections was the water ice crystalline \citep{terada_tokunaga12, schegerer_wolf10}.  This result is consistent with protostellar collapse models, which predict that the bulk of the disk ice in the upper layers of the outer disk arrives unprocessed from its initial chemical formation on grains in the parent molecular cloud or protostellar envelope \citep{visser+09a}, so it should be amorphous. Significantly, in one of the two disks with previous crystalline ice detections, YLW16A, the disk and envelope water ice were spatially resolved, with crystalline ice only in the upper layers of the outer disk and amorphous ice in the inner disk and in falling envelope \citep{schegerer_wolf10}.

In contrast, near- and far-infrared spectral observations of solar system objects find compositions dominated by crystalline ice. \citet{jewitt_luu04} detected a 1.65$\mu$m crystalline ice signature towards Kuiper belt object (KBO) Quaoar, while \citet{grundy+06} found the same feature in satellites of Uranus, placing an upper limit of 20\% on the amorphous content of the observed water ice.  At longer wavelengths, ISO SWS/LWS spectra of comet Hale-Bopp show both crystalline features at 43 and 63$\mu$m, which are best-fit with 15$\mu$m sized ice grains \citep{lellouch+98}, an order of magnitude greater than the size inferred in edge-on disk features.  It is possible to crystalize amorphous ice on large bodies through collisions of planetesimals or differentiation \citep{brown12}.

Depending on the initial phase of ice in the disk, there are processes by which it can be amorphized or crystallized.  Ice which is initially crystalline can recondense amorphously after being photodesorbed in the upper layers of the outer disk \citep{ciesla14a}.  Alternatively, its crystalline structure can be damaged by high-energy radiation or particles when the disk temperature is less than 80K \citep[and references therein]{grundy+99}, which should amorphize these regions completely on the order of 10$^5$-10$^6$ years \citep{cook+07}.  On the other hand, local heating events, such as shocks or collisions with other dust grains or large bodies, can crystallize amorphous ice \citep{porter+10, marboeuf+09} and ice can be crystallized as material moves into warmer regions near the star.  

Our finding of amorphous ice in the upper layers of AA Tau's outer disk is consistent with the prediction that ice below 80K should become amorphous by 10$^6$ years, if it was not already inherited that way from its natal cloud.  The region where crystalline ice could be thermally generated is much smaller than the contribution region of the 47$\mu$m amorphous feature. However, the case of DO Tau is more complicated.  The vast majority of its outer disk is below 80 K and should, therefore, be amorphous.  The remaining regions above 80K are almost entirely within the complete photodesorption zone, so crystalline ice should not exist in the upper layers of its disk.  Our detection of crystalline ice in this system suggests that there has been replenishment of the crystalline ice population in the outer disk at some point within 10$^5$ years.  Replenishment by warm ice from the inner disk would require transport through the disk.  However, in the disk's upper layers this would require the crystalline ice to pass through regions where turbulence should circulate grains into the photodesorption zone, which would have the effect of amorphizing the grains \citep{ciesla14a}.  Therefore our detection strongly suggests regeneration by in situ crystallization of icy grains, e.g. by micrometeorite/planetesimals collisions or a desorption/recondensation event. We note that these results confirm the conclusions of \citet{mcclure+12}.

\subsection{Can we trace the radial extent of the water ice snowline?}
\label{iceset}

As seen by DO Tau, photodesorption can have a significant effect on the curvature of the snow line in the upper layers, releasing water ice from grains at larger radii than thermal desorption alone. Our attempts to fit DO Tau with a radially constant disk model suggest that inclusion of ice inside this region in our disk structure models (which do not self-consistently account for photodesorption) causes the decrease in flux from 20 to 30$\mu$m, which prevents us from fitting the IRS and PACS spectra simultaneously.  We confirmed the general effect on that wavelength region of truncating radially the water ice in the disk upper layers inwards of some critical radius $R_C$ in Figure \ref{dotau_rcs}, using the new 2-zone model.  Although DO Tau was not fit perfectly with the new model prescription, the broader mid- to far-IR SED fit is improved by models with less ice in the inner 100AU. Including a curved desorption zone rather than a step function may improve the fit. 

These results suggest that with updated modeling, including a self-consistent account of photodesorption, it may be possible to constrain the location of water ice in the upper layers of the inner $\sim$50AU by comparing the strength of the 63$\mu$m feature with the absolute flux and slope of the SED over the end of the {\it Spitzer} IRS spectrum.  Combining this type of analysis with a more physical settling parameterization, and potentially data around 45$\mu$m with FIFI on SOFIA, SPICA, or near-infrared scattered light spectroscopy (e.g. with GPI or SPHERE) may allow a more complete picture of the 2D location water ice depletion front, or `snowline', taking into account both thermal and photo-chemical effects.  

\section{Conclusions}
\label{conclusions}

We present four new {\it Herschel} PACS spectroscopic observations of disks around T Tauri stars in the Taurus and Cha I star-forming regions.  Two of the Taurus disks show crystalline water ice features at 63$\mu$m, while we infer the presence of the red wing of the amorphous 47$\mu$m water ice feature in the third system.  A fourth disk, in Cha I, exhibits an ice non-detection and is used as a control.  

Using detailed irradiated accretion disk models, we extract basic constraints on the abundance, grain size, and radial and vertical location of the water ice in the disk, finding that:

\begin{itemize}
\item both of the amorphous 47 and crystalline 63$\mu$m features are dominated by emission at trans-Neptunian radii, R$>$30AU, in the disk upper layers;
\item the emitting region of the crystalline ice is much larger than the region that is hotter than the crystallization temperature, suggesting local heating or transport; 
\item both features were well-fit by an ice mass fraction of 0.002 relative to gas, or half the predicted solar nebula value, consistent with a depletion of ice from the disk upper layers;
\item comparing the ice feature strength with continuum shape in {\it Spitzer} IRS may yield more detailed information regarding the location of the snowline in the upper layers.
\end{itemize} 

Through this work, we probe the main reservoir thought to provide water ice to terrestrial planets, namely the proto-Kuiper Belt.  However, larger disk sample sizes and more physically motivated model ice distributions are necessary to characterize better the properties of this reservoir and determine the innermost spatial extent of water ice in disks, i.e. the snowline, both of which are essential to constraining how Earth acquired its water.

\acknowledgments
M.K.M was supported by the National Science Foundation Graduate Student Research Fellowship under Grant No. DGE 0718128.  N.C. acknowledges support from NASA Origins grants NNX08AH94G.

\clearpage



\clearpage

\clearpage

\clearpage

\begin{deluxetable}{ccccc}
\tabletypesize{\small}   
\tablewidth{0pt}
\tablecaption{Stellar and Model Properties}
\tablehead{\colhead{Parameter} & \colhead{AA Tau} & \colhead{Haro 6-13$^a$} & \colhead{DO Tau} & \colhead{VW Cha}}
\startdata
 \hline \\
$T_{eff}$ (K) 					& 4060 & 3850  & 3850  &  4350 \\
$A_V$ (mag)     				& 1.34     & 7.0    & 3.0  &   2.8 \\
$M_*$  (\Msun)    				& 0.80     &  0.56 & 0.56  & 1.1  \\
$R_*$ (\Rsun) 					& 1.80     &  2.45 &  1.90 &  2.7 \\
$\dot{M}$ (\Msun/yr) 				&  $6\times10^{-9}$  & $1\times10^{-8}$   & $9\times10^{-8}$  & $5\times10^{-8}$  \\
$i$ (\degr)   					& 71  & 40  & 42  & 45$^b$  \\
$d$  (pc) $^{c}$   				& 140  & 140  & 140  & 160  \\
 \hline \\
Wall, lower \\
\hline \\
$T_{wall,1}$ (K) 		  		&  1600 		   &  1400   		  & 1600 &  1600 \\
$a_{max}$ ($\mu$m) 	  		&  1			    &  5   			  & 2 & 3  \\
sil. comp.                                                & 100\% PyMg60 &  100\% PyMg60   & 100\% PyMg60 & 100\% PyMg60 \\
$R_{wall,1}$ (AU)          			&       0.12  	    &   0.15  & 0.17 & 0.21 \\
$h_{wall,1}^5$ = $z_{wall,1}$ (AU)       & 	0.009 (2.25H)	    &   0.024 (4H)  & 0.021 (3H) &  0.018 (2.5H) \\
$z_{s,disk} (R_{wall})$ (AU) 		&	9.7$\times10^{-3}$   &	1.6$\times10^{-2}$  &  2.1$\times10^{-2}$ & $2.4\times10^{-2}$  \\
\hline \\
Wall, upper \\
\hline \\
$T_{wall,2}$ (K)             			&     750  		   & 800	  & 700 &  1200 \\
$a_{max}$ ($\mu$m) 			&     5	            &  3  & 3 &   0.75 \\
$sil.$ $comp.$ 					& 60\% OlMg50  &  100\% PyMg60   & 60\% OlMg50  &  50\% OlMg50 \\
							& 40\% Forst.	   &            & 40\% Forst. &  50\% Forst. \\
$R_{wall,2}$ (AU)    				&  	0.32		   &  0.36   & 0.57  & 0.35  \\
$h_{wall,2}^d$ (AU)				&   	0.035 (3H)	   &   0.05(3.5H)  &  0.07 (2.75H)  &  0.012 (1H) \\
$z_{wall,2}$ (AU)          			&      	0.044	   &   0.074  & 0.085  & 0.030  \\
$z_{s,disk} (R_{2})$ (AU) 			&    	3.4$\times10^{-2}$	   &   5.3$\times10^{-2}$  & 9.1$\times10^{-2}$ &  $4.8\times10^{-2}$ \\
 \hline \\
 Disk &  &   &   &   \\
 \hline  \\
$\alpha$       					& 0.003 				& 0.005  & 0.05  &  0.005-0.01$^e$ \\
 $\epsilon$   					& 0.01   				& 0.1  	   & 0.05  &  0.1-0.5$^e$ \\
 $R_{d}$ (AU)      				& 140    				& 180      & 350   &  7 \\
 $M_{d}$ (\Msun)     				&  3.06$\times$10$^{-2}$  &   3.67$\times10^{-2}$  &   4.06$\times10^{-2}$  & 6.69$\times$10$^{-3}$  \\
 $a_{max,midplane}$$^f$		&  1mm &  1mm   & 1cm   & 1mm  \\ 
 silicates/graphite$^g$	    		&   &   &   &   \\
 $a_{max, upper}$ ($\mu$m)  		& 0.25  		      & 0.25   		& 0.25  	& 0.5  \\
 sil. species					&  100\% PyMg80  & 90\% OlMg50   &   100\% OlMg50		&  50\% PyMg80 \\
							&			      &	 10\% Enst.	&    		&  30\% Forst. \\
							&			      &			         &    		&  20\% Enst. \\
  H$_2$O ice    					&  			      &          			&   		&   \\
 $a_{max, upper}$ ($\mu$m) 		& 0.25 & 15  & 0.25  &  0.25-15 \\
 $m/m_{H_2}$ 					& 2$\times$10$^{-3}$ & 2$\times$10$^{-3}$  & 2$\times$10$^{-3}$  & up to 2$\times$10$^{-3}$  \\
 $power$    					& 2.0 &   & 3.5  &   2.0
\enddata
\tablecomments{
\\
$^a$ Our model for Haro 6-13 has a gap between 0.36AU and 7.5AU, with an outer wall of 4$H$ and a composition the same as that of the disk. \\
$^b$ VW Cha lacks definitive inclination information, so we assumed $i$=45\degr.\\
$^c$ Distance references: Taurus: \citet{kenyon+94},  Chamaeleon I: \citet{whittet+97}, references for stellar and accretion parameters are given in \S\ref{input}.\\
$^d$ The wall height is expressed in terms of the gas pressure scale height, $H$, at the wall radius.  \\
$^e$ VW Cha lacks data beyond $\sim$200$\mu$m; without that additional constraint, $\alpha$ and $\epsilon$ can vary as indicated. \\
$^f$ The maximum grain size in the midplane was taken to be the same for silicates, graphite, and water ice, which are assumed to be present in the midplane with the same abundance as in the upper layers. \\
$^g$ Silicates and graphite were assumed to have a power of -3.5 and $m/m_{H_2}$=4$\times$10$^{-3}$ and 2.5$\times$10$^{-3}$, respectively. \\
 }
\label{modelfits}
\end{deluxetable}

\end{document}